\documentclass[useAMS, usenatbib]{mn2e}
\usepackage{amssymb,amsmath}
\usepackage{graphicx}
\usepackage{xcolor}
\usepackage{multirow, multicol}

\def\oiii{[O~{\sc iii}]\ }
\def\nii{[N~{\sc ii}]\ }

\title[{\rm \oiii} variability]
{Apparent \oiii variability in the narrow line Seyfert I Mrk142}
\author[Zhang \& Feng]
       {Xue-Guang Zhang\thanks{zhangxg23@sysu.edu.cn} \& 
        Long-Long Feng\\ 
       Institute of Astronomy and Space Science, Sun Yat-Sen University, 
          No. 135, Xingang Xi Road, Guangzhou, 510275, P. R. China}

\date{}
\def\LaTeX{L\kern-.36em\raise.3ex\hbox{a}\kern-.15em
    T\kern-.1667em\lower.7ex\hbox{E}\kern-.125emX}

\begin{document}
\pagerange{\pageref{firstpage}--\pageref{lastpage}} \pubyear{2015}
\maketitle
\label{firstpage}

\begin{abstract}
    In this letter, we checked spectral properties of the 
well-known narrow line Seyfert I Mrk142, in order to try to find effects 
of narrow line variability on BLR radius of Mrk142 which is 
an outlier in the R-L plane. Although, no improvement can be found on 
BLR radius, apparent narrow line variability can be confirmed in Mrk142. 
Using the public spectra collected from the Lick AGN Monitoring Project, 
the spectral scaling method based on assumption of constant \oiii line 
is firstly checked by examining broad and narrow emission line properties. 
We find that with the application of the spectral scaling method, there is 
a strong correlation between the \oiii line flux and the \oiii line 
width, but weaker correlations between the broad H$\alpha$ flux and 
the broad H$\beta$ flux, and between the broad H$\alpha$ flux and the 
continuum emission at 5100\AA. The results indicate that the assumption 
of constant \oiii line is not preferred, and caution should be exercised 
when applying the spectral scaling calibration method. And then, we can 
find a strong correlation between the \oiii line flux and the continuum 
emission at 5100\AA, which indicates apparent short-term variability of 
the \oiii line in Mrk142 over about two months.
\end{abstract}


\section{Introduction}
   
    Emission line variability is a powerful tool to study structures of 
central emission line regions of active galactic nuclei (AGN). Broad 
emission line variability has been studied to determine structures of 
central broad emission line regions (BLR) through the well-known 
reverberation mapping technique (Blandford \& Mckee 1982, Peterson et al. 
1993, Bentz et al. 2010, Zhang 2011, Grier et al. 2013, Zhang 2013, 
Pancoast et al. 2014, Zhang 2015). However, similar methods on narrow line 
variability is hard to apply to study central narrow emission line 
regions (NLR) of AGN, because we do not see expected short-term narrow 
line variability of AGN. Furthermore, when we study broad line variability 
of AGN, the assumption of a constant narrow line (commonly the \oiii line) 
is commonly accepted to do the spectral scaling algorithm (Van Groningen \& 
Wanders 1992) to normalize flux calibration of the spectra to a consistent 
scale for spectra observed with different configurations in 
different observatories (Peterson et al. 1998, Bentz et al. 2009, and also 
see references for the other mapped AGN).

   However, some results have been reported on narrow line variability.  
Zheng et al. (1995) reported ten-year narrow line variability in 3c390.3. 
And, Peterson et al (2013) reported properties of long-term narrow line 
variability in NGC5548. If there was true short-term narrow line variability 
(for example, narrow lines from emission regions with high electron density 
around $10^6cm^{-3}$, Peterson et al. 2013), the commonly applied spectral 
scaling calibration method based on constant \oiii line should lead to 
some unreasonable results. Here, in this letter, based on collected spectra 
with and without applications of the spectral scaling calibration method, 
we carefully checked spectral properties of the well-studied mapped AGN 
Mrk142, and found that the spectral scaling calibration method could be 
unreasonable, and then report more interesting results on narrow line 
variability. This letter is organized as follows. Section 2 shows our main 
results and necessary discussions, and then Section 3 gives our conclusions.
 
\section{Main Results}

\begin{figure*}
\centering\includegraphics[width = 18cm,height=3.5cm]{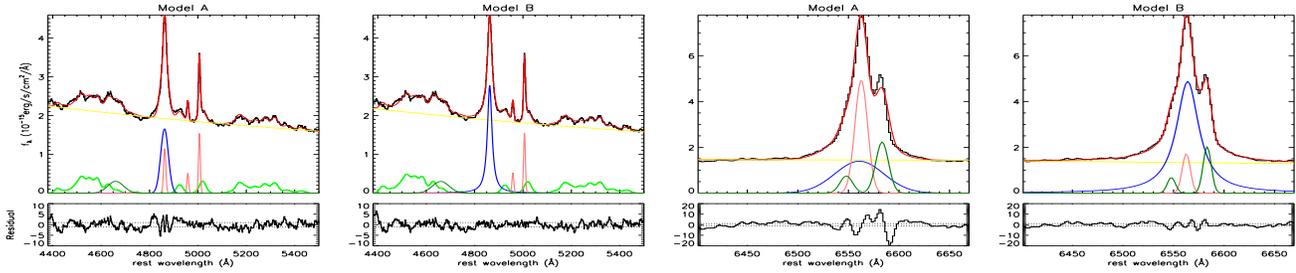}
\caption{Examples of the best fitted results for emission lines in the 
final spectrum observed in 12th, April 2008, by different model functions. 
The left two panels and the right two panels show the results based on 
Model A and Model B functions for the emission lines around H$\beta$ and 
around H$\alpha$ respectively. Top and bottom panels show the best fitted
results and the corresponding residuals respectively. In each top panel,
solid lines in black, in red, in yellow and in blue show the spectrum,
the best fitted results, the power law AGN continuum emission and the
determined broad balmer lines respectively. And in the left two top panels,
solid lines in green, in dark green and in pink show the Fe~{\sc ii}
components, the He~{\sc ii} line and the narrow H$\beta$ and the \oiii
doublet respectively. In the right two top panels, solid lines in dark
green and in pink show the \nii doublet and the narrow H$\alpha$
respectively. In each bottom panel, the horizontal dotted lines
are for $residual=\pm1$.
}
\label{line}
\end{figure*}

    Mrk142 is a well-studied mapped AGN in the Lick AGN monitoring 
Project (LAMP, http://www.physics.uci.edu/\~{}barth/lamp.html). The main 
reason to select Mrk142 as the target of this letter is that Mrk142 is the 
unique outlier in the Radius-Luminosity (R-L) plane for the mapped nearby 
broad line AGN (Bentz et al. 2013). 

   Here, we collect the two kinds of spectra of Mrk142 from the LAMP data 
release: the final reduced spectra (hereafter, 'final spectra') and the 
scaled spectra after the application of the spectral scaling method 
(hereafter, 'scaled spectra'). The 'final spectra' are the directly reduced 
spectroscopic data with flux calibrations determined from nightly 
spectra of standard stars (typically Feige 34 and BD+284211), which have 
been well done by the LAMP (more detailed descriptions on the spectroscopic 
technique can be found in Bentz et al. 2009). Under the assumption of a 
constant \oiii line, the 'scaled spectra' are obtained by applying the 
spectral scaling method to the 'final spectra', in order to mitigate the 
effects of different observational configurations. Then, we can carefully 
check the emission lines in the spectra of Mrk142.

    Before proceeding further, there are three points we should note. 
First and foremost, as discussed in Park et al. (2012) and Hu et al. 
(2015), an additional star component could be included in the spectra of 
Mrk142. However, when we try to add a star component to describe the 
LAMP spectrum of Mrk142, it is hard to find a clear and stable star 
component. And moreover, even without a star component, the best 
description of the emission lines can  be achieved. Therefore, we 
do not consider an additional star component in this letter any more. 
Different model functions can be applied to describe emission lines, such 
as gaussian functions, lorentz functions, gauss-hermite functions, etc.. 
Actually, for the emission lines of Mrk142, simple gaussian or lorentz 
functions are good enough to describe them, which can be assured by 
the results shown in the following Figure~\ref{line}. Therefore, we do 
not discuss more complex model functions for the emission lines. Last 
but not least, we should note that there are apparent effects of slit 
losses, variable seeing and transparency etc. on the measured line 
parameters from the final spectra. In order to mitigate the effects, 
the spectral scaling method is commonly applied. Therefore, if the 
spectral scaling method was preferred, we would expect stronger parameter 
correlations from the scaled spectra than from the final spectra, because 
problems with the final spectra have been corrected as much as possible. 
Hence, we mainly focus on the comparisons between the results from the 
final spectra and from the scaled spectra. 

   Now, the emission lines can be described as follows for the emission 
lines around H$\beta$ (Fe~{\sc ii}, He~{\sc ii}, H$\beta$ and \oiii doublet) 
and around H$\alpha$ (H$\alpha$ and \nii doublet). The Fe~{\sc ii} template 
in Kovacevic et al. (2010) is applied to describe the optical Fe~{\sc ii} 
lines, a power law function is applied to describe the AGN continuum 
emission, a broad gaussian function is applied to describe the 
He~{\sc ii} line, two narrow gaussian functions are applied to describe 
the \oiii (the \nii) doublet. And, when the \oiii and \nii doublets are 
described by the gaussian functions, their flux ratios are fixed to the 
theoretical values ($f_{5007}/f_{4959}=f_{6585}/f_{6549}=3$). For the 
broad and narrow balmer lines, two different model functions 
applied, Model A: one broad plus one narrow gaussian functions, Model B: 
one broad lorentz function plus one narrow gaussian function. For the 
model functions applied to the narrow balmer lines, they have the 
same line widths. Here, a very weak narrow He~{\sc ii} line is not 
considered. Then, through the Levenberg-Marquardt least-squares 
minimization technique, the emission lines can be well fitted. 

    Figure~\ref{line} shows examples of the best fitted results for the 
emission lines by the model functions. We can clearly see that the 
lorentz functions for the broad balmer lines can lead to better 
fitted results, due to smaller residuals around H$\beta$ and H$\alpha$. 
Here, the residuals are calculated by $\frac{y_{obs}-y_{fit}}{y_{err}}$, 
where $y_{obs}$, $y_{err}$ and $y_{fit}$ represent the observed values, 
the corresponding uncertainties and the best fitted results respectively. 
Therefore, in this letter, only Model B is considered, and no complex 
model functions are discussed. Then, we carefully check the line parameters 
determined from Model B as follows.

\begin{figure}
\centering\includegraphics[width = 7cm,height=4.5cm]{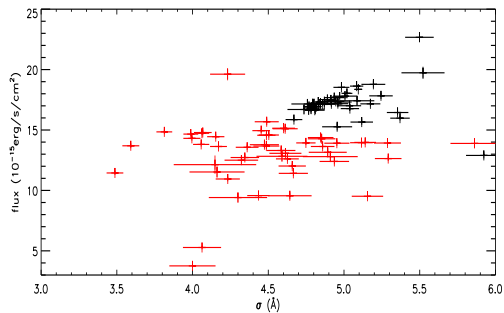}
\caption{On the correlations between line flux and line width
of the \oiii lines in the scaled spectra (symbols in black) and in the
final spectra (symbols in red) respectively.
}
\label{scale_o3}
\end{figure}

   First and foremost, we checked line parameter correlations of the \oiii 
line. If the spectral scaling method was preferred, we would expect no line 
parameter correlation to be found for the \oiii lines in the scaled spectra of 
Mrk142. Figure~\ref{scale_o3} shows the correlations between line flux 
and line width of the \oiii lines in the scaled spectra and in the final 
spectra respectively. In the scaled spectra, there is a apparent 
linear correlation with spearman rank correlation coefficient about 0.49 
with $P_{null}\sim10^{-4}$. But, there is no clear linear correlation 
from the final spectra. Therefore, by the correlations shown in 
Figure~\ref{scale_o3}, the applied spectral scaling calibration method 
is not preferred. 

\begin{figure}
\centering\includegraphics[width = 7cm,height=4.5cm]{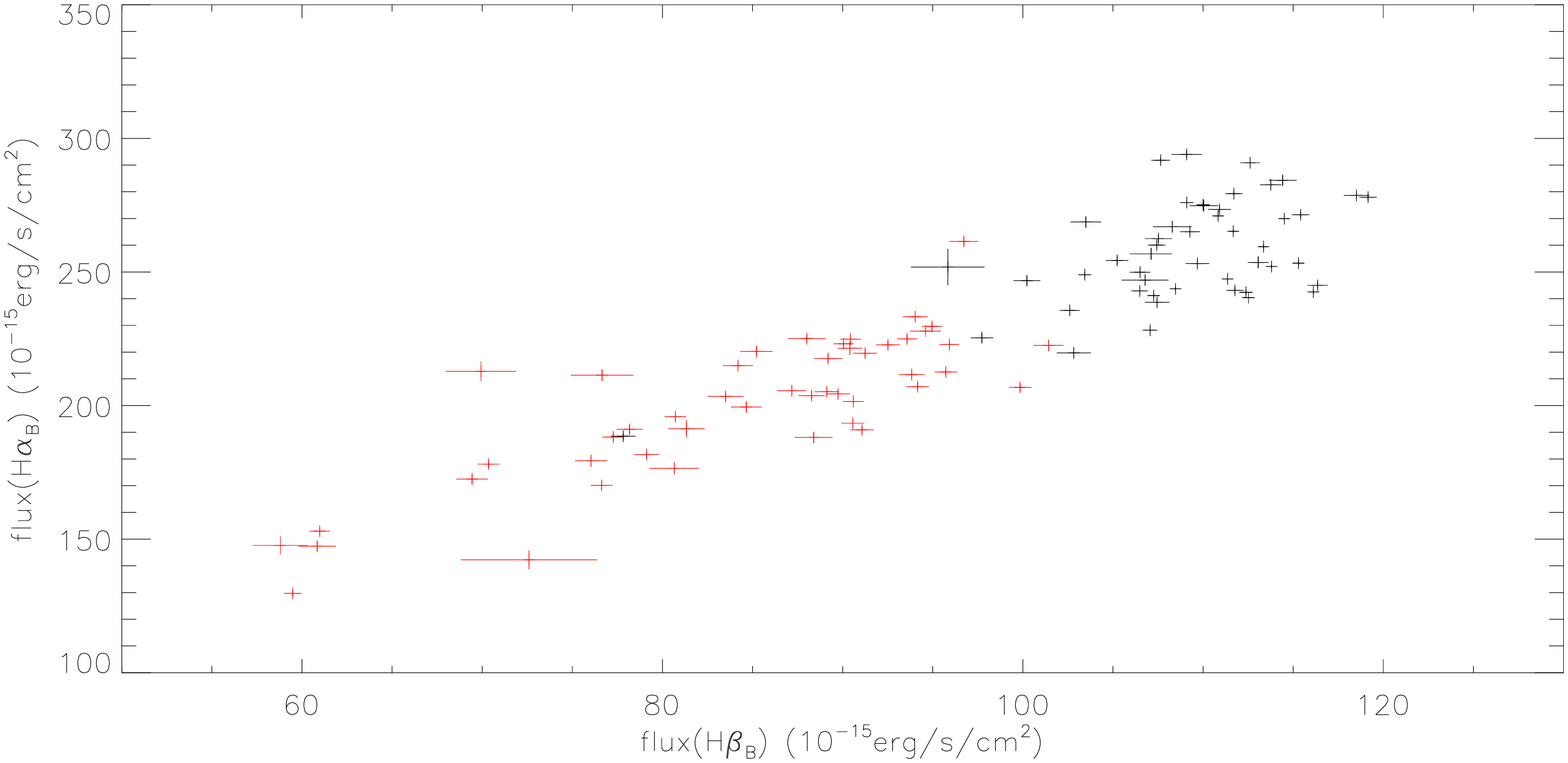}
\caption{On line flux correlations between the broad balmer lines
in the scaled spectra (symbols in black) and in the final spectra
(symbols in red) respectively.
}
\label{hab}
\end{figure}

    Besides, we checked flux correlations between the broad balmer lines. 
If the application of the spectral scaling calibration method were 
reasonable, a stronger line flux correlation could be expected from 
the scaled spectra than from the final spectra. Figure~\ref{hab} shows 
the broad line flux correlations. The spearman rank correlation 
coefficients are about 0.44 with $P_{null}\sim10^{-3}$ and 0.79 with 
$P_{null}\sim10^{-11}$ for the broad balmer lines in the scaled spectra 
and in the final spectra respectively. Therefore, the results in 
Figure~\ref{hab} also indicate that the spectral scaling calibration 
method is not preferred.

\begin{figure}
\centering\includegraphics[width = 7cm,height=4.5cm]{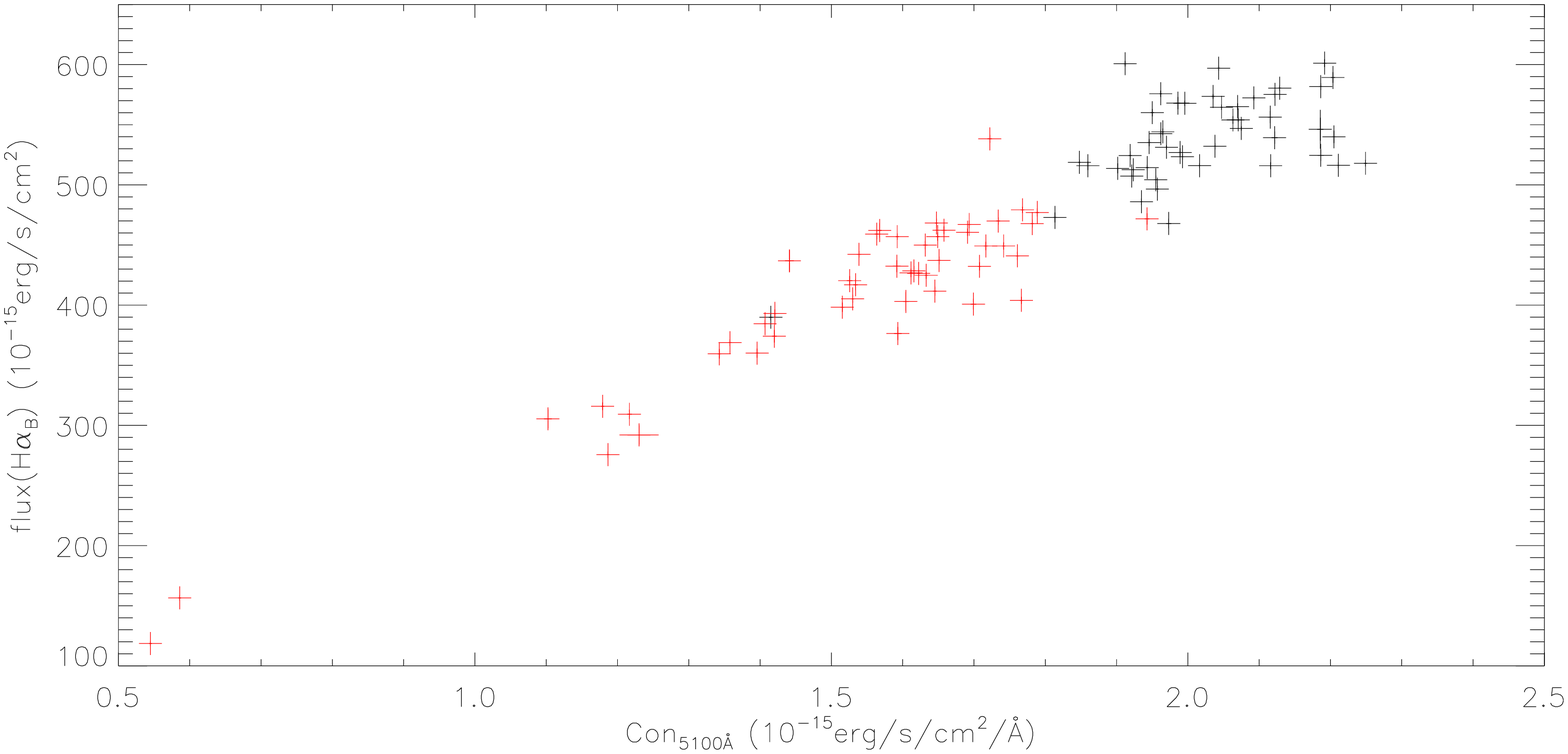}
\caption{On the correlations between broad line flux of H$\alpha$ and 
continuum emission at 5100\AA\ from the scaled spectra (symbols in black) 
and from the final spectra (symbols in red) respectively. 
}
\label{con}
\end{figure}

    Last but not least, we checked the correlations between continuum 
emission at 5100\AA\ and broad line flux of H$\alpha$ in the scaled  
spectra and in the final spectra respectively, which are shown in the 
Figure~\ref{con}. Here, similar to what was done by Greene \& Ho 
(2005), the line flux of broad H$\alpha$ includes the contributions from 
the \nii doublet and the narrow H$\alpha$. The spearman rank 
correlation coefficients are about 0.49 with $P_{null}\sim10^{-4}$ and 
0.81 with $P_{null}\sim10^{-12}$ for the parameters from the scaled  
spectra and from the final spectra respectively. Therefore, the results 
shown in Figure~\ref{con} also indicate that the spectral scaling calibration 
method is not preferred, otherwise a stronger linear correlation 
could be expected from the scaled spectra.

   Based on the results above, after careful consideration, we concluded 
that the assumption of constant \oiii lines to do the spectral scaling 
calibration is probably not reasonable for Mrk142, which indicates 
possible \oiii variability to some extent. Then, we can show one 
rough but interesting check for the \oiii line variability in Mrk142 
through the collected final spectra. Certainly, the absolute flux 
calibration of the final spectra is based on the spectra of standard stars, 
which tends to give more reliable results on line flux than on line width. 
However, we should note that it is hard to correct the effects of slit 
losses, variable seeing and transparency etc. on the line parameters from 
the final spectra, once we have accepted that the spectral scaling method is 
not preferred. Therefore, we still show the comparisons on the \oiii line 
flux to check variability of the \oiii line.

\begin{figure}
\centering\includegraphics[width = 7cm,height=9cm]{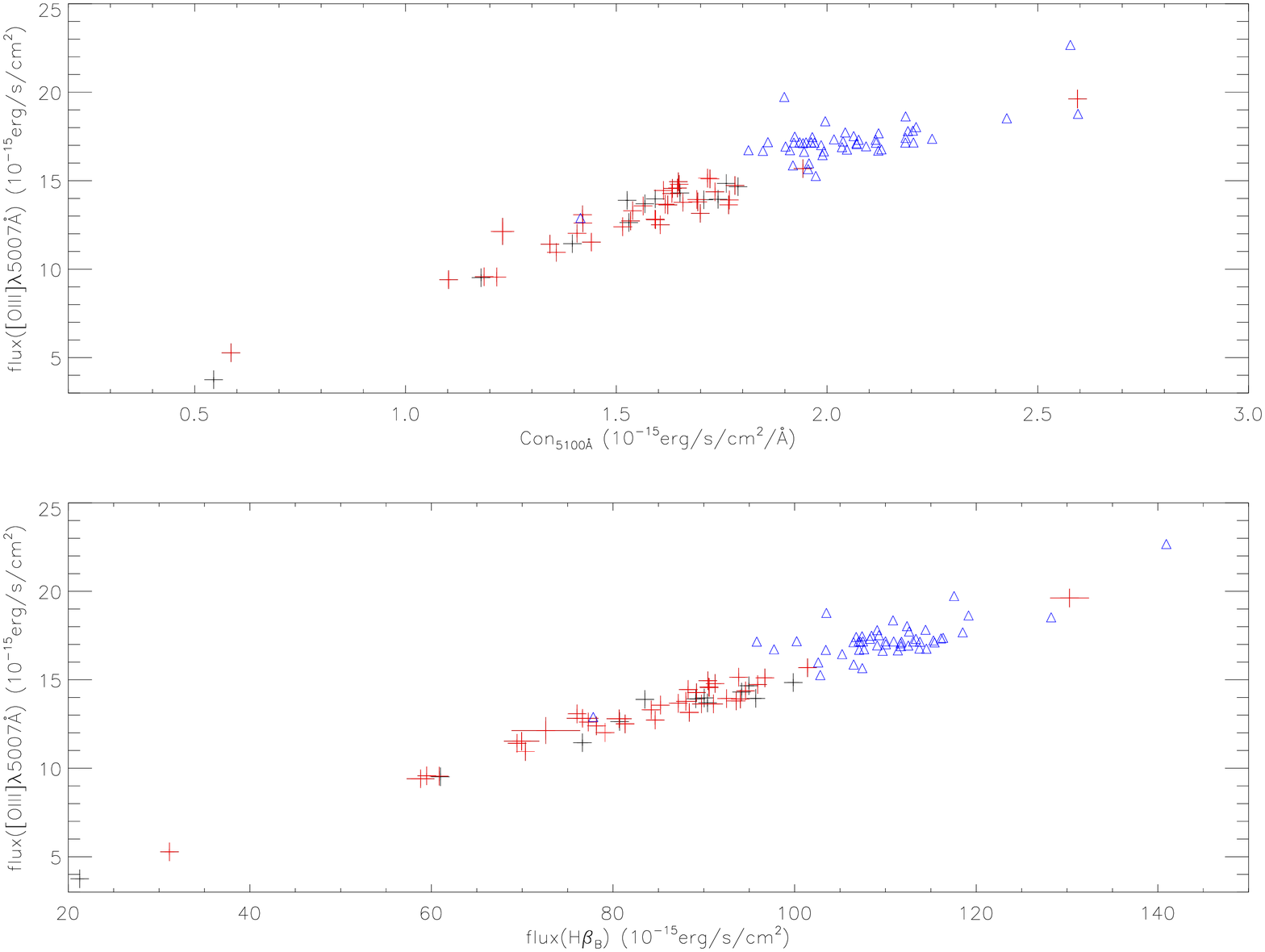}
\caption{On the correlation between \oiii line flux and continuum
emission at 5100\AA\ (top panel), and the correlation between \oiii line 
flux and broad H$\beta$ line flux (bottom panel), using the final 
spectra (symbol of plus) and using the scaled spectra (symbol of 
triangle). In each panel, plus symbols in black represent the values 
from the final spectra with smaller or larger \oiii line widths.
}
\label{var_o3}
\end{figure}

    Figure~\ref{var_o3} shows the correlation between \oiii line 
flux and continuum emission at 5100\AA (broad H$\beta$ line flux). There 
are strong linear correlations with spearman rank correlation coefficients 
of about 0.87 with $P_{null}\sim10^{-17}$ and 0.93 with $P_{null}\sim10^{-22}$ 
for the correlations on continuum emission and on broad H$\beta$ 
line flux respectively from the final spectra. The strong correlations 
indicate there is reliable short-term \oiii variability over about two 
months. Furthermore, we should note that in some final spectra, the 
measured \oiii line widths are smaller or larger than the mean value 
(that is another reason why it is necessary to do the spectral scaling 
calibration). Even the spectra with smaller or larger \oiii line widths 
are rejected by $|\sigma-\bar{\sigma}|>0.5\AA$, the correlations are 
still strong with coefficients about 0.89 with $P_{null}\sim10^{-14}$ and 
0.93 with $P_{null}\sim10^{-18}$ respectively. Furthermore, from the 
scaled spectra, the linear correlation between \oiii line flux and 
continuum emission at 5100\AA\ (broad H$\beta$ line flux) can also 
be found, although there is a small variability amplitude in the 
\oiii line flux. The spearman rank correlation coefficients are about 
0.51 with $P_{null}\sim10^{-4}$ and 0.48 with $P_{null}\sim10^{-4}$ for 
the correlations on continuum emission and on broad H$\beta$ line flux 
respectively for the parameters from the scaled spectra. It is clear 
that the results shown in Figure~\ref{var_o3} can not be explained by 
the effects of slit losses, variable seeings and transparency etc., 
because of no dependence of random slit losses on central continuum 
emission (broad balmer line emission) and because of such tiny variability 
of seeings and air masses during the observational period for Mrk142.

    Before the end of the subsection, there are three points we should 
discuss further. First and foremost, we do not consider the effects 
of probable contributions of star components. Actually, our measured 
line parameters (especially for the \oiii line) do not depend  
on whether there are contributions from star components. Therefore, the 
results above on the line parameters can be well accepted. Besides, we 
can be sure that the spectral scaling calibration method is not preferred 
for the spectral of Mrk142. However, no improved results on BLR radius 
can be found on cross-correlation of the variability of continuum emission 
and broad line emission (time lag from the cross-correlation function is 
still near zero), even with considerations of the \oiii variability. 
And moreover, due to expected \oiii variability, it is hard to correct 
the effects of slit losses, variable seeings and transparency etc. on 
the line parameters from the final spectra. In other words, we can not 
give a clearer conclusion on the \oiii variability with a reliable 
variability amplitude, but we can find strong evidence to support the 
\oiii variability. Last but not least, we also tried to check the spectral 
properties of several another mapped AGN by the same procedures above, but we 
can not find similar results as those of Mrk142. In other words, the spectral 
scaling calibration method with the assumption of a constant \oiii line could 
be commonly accepted to some extent, but caution should be exercised 
when applying the calibration method. And checking corresponding 
correlations on spectral parameters in the spectra with and without 
the application of a calibration method could provide further information on 
whether the calibration method is preferred.

\begin{table*}
\begin{minipage}{17cm}
\caption{Line parameters}
\begin{tabular}{llllll|lllll}
\hline
HJD & \multicolumn{5}{c}{scaled spectra} & \multicolumn{5}{c}{final spectra}\\
  & con & $f_\beta$ & $f_\alpha$ & $\sigma_3$ & $f_3$ &
    con & $f_\beta$ & $f_\alpha$ & $\sigma_3$ & $f_3$ \\
\hline
4550  &  2.2$\pm$0.1  &  112.3$\pm$0.3  &  242.4$\pm$0.9  &  5.0$\pm$0.1  &  18.1$\pm$0.1  &  1.1$\pm$0.1  &  59.4$\pm$0.5  &  129.6$\pm$1.1  &  4.4$\pm$0.1  &  9.5$\pm$0.1  \\
4551  &  1.9$\pm$0.1  &  102.5$\pm$0.5  &  235.5$\pm$1.2  &  5.3$\pm$0.1  &  15.2$\pm$0.1  &  1.7$\pm$0.1  &  89.1$\pm$0.7  &  205.1$\pm$1.4  &  5.2$\pm$0.1  &  13.1$\pm$0.2  \\
4553  &  2.0$\pm$0.1  &  113.3$\pm$0.2  &  259.4$\pm$0.7  &  4.8$\pm$0.1  &  17.2$\pm$0.1  &  1.7$\pm$0.1  &  95.9$\pm$0.5  &  222.8$\pm$1.4  &  4.0$\pm$0.1  &  14.3$\pm$0.1  \\
4555  &  2.5$\pm$0.1  &  140.9$\pm$1.5  &  325.8$\pm$4.3  &  5.4$\pm$0.1  &  22.1$\pm$0.3  &  1.1$\pm$0.1  &  58.8$\pm$1.5  &  147.6$\pm$3.4  &  4.2$\pm$0.1  &  9.4$\pm$0.4  \\
4556  &  2.1$\pm$0.1  &  111.7$\pm$0.4  &  243.0$\pm$1.2  &  4.8$\pm$0.1  &  17.4$\pm$0.1  &  1.9$\pm$0.1  &  101.4$\pm$0.8  &  222.5$\pm$1.9  &  4.4$\pm$0.1  &  15.$\pm$0.2  \\
4557  &  1.9$\pm$0.1  &  110.9$\pm$0.6  &  273.4$\pm$1.2  &  5.1$\pm$0.1  &  17.3$\pm$0.2  &  1.5$\pm$0.1  &  90.1$\pm$0.5  &  223.1$\pm$1.1  &  5.1$\pm$0.1  &  13.1$\pm$0.1  \\
4558  &  2.1$\pm$0.1  &  118.5$\pm$0.7  &  278.6$\pm$1.2  &  4.9$\pm$0.1  &  17.0$\pm$0.2  &  1.7$\pm$0.1  &  94.6$\pm$0.8  &  227.8$\pm$1.3  &  4.8$\pm$0.1  &  14.0$\pm$0.2  \\
4559  &  1.9$\pm$0.1  &  107.4$\pm$0.7  &  238.6$\pm$1.4  &  5.1$\pm$0.1  &  15.2$\pm$0.2  &  1.7$\pm$0.1  &  95.7$\pm$0.6  &  212.5$\pm$1.3  &  5.1$\pm$0.1  &  13.2$\pm$0.2  \\
4560  &  2.0$\pm$0.1  &  115.4$\pm$0.5  &  271.4$\pm$1.1  &  4.8$\pm$0.1  &  17.2$\pm$0.1  &  1.6$\pm$0.1  &  92.5$\pm$0.6  &  222.7$\pm$1.3  &  4.7$\pm$0.1  &  13.0$\pm$0.1  \\
4561  &  1.9$\pm$0.1  &  105.2$\pm$0.6  &  254.3$\pm$1.5  &  5.3$\pm$0.1  &  16.1$\pm$0.2  &  1.5$\pm$0.1  &  80.7$\pm$0.6  &  195.8$\pm$1.4  &  5.2$\pm$0.1  &  12.1$\pm$0.2  \\
4562  &  2.2$\pm$0.1  &  114.4$\pm$0.7  &  284.3$\pm$1.8  &  5.2$\pm$0.1  &  17.1$\pm$0.2  &  1.1$\pm$0.1  &  60.9$\pm$0.5  &  152.9$\pm$1.3  &  5.1$\pm$0.1  &  9.5$\pm$0.1  \\
4564  &  2.0$\pm$0.1  &  111.6$\pm$0.2  &  265.2$\pm$0.8  &  4.7$\pm$0.1  &  17.2$\pm$0.1  &  1.7$\pm$0.1  &  94.9$\pm$0.5  &  229.6$\pm$1.3  &  3.9$\pm$0.1  &  14.1$\pm$0.1  \\
4566  &  2.1$\pm$0.1  &  113.7$\pm$0.6  &  282.6$\pm$0.9  &  5.0$\pm$0.1  &  16.2$\pm$0.1  &  1.7$\pm$0.1  &  94.0$\pm$0.6  &  233.2$\pm$1.1  &  4.9$\pm$0.1  &  13.2$\pm$0.2  \\
4567  &  2.0$\pm$0.1  &  115.2$\pm$0.3  &  253.2$\pm$1.1  &  4.8$\pm$0.1  &  17.2$\pm$0.1  &  1.6$\pm$0.1  &  89.7$\pm$0.6  &  204.3$\pm$1.5  &  4.1$\pm$0.1  &  13.1$\pm$0.1  \\
4568  &  2.0$\pm$0.1  &  114.5$\pm$0.3  &  269.9$\pm$0.9  &  4.7$\pm$0.1  &  16.2$\pm$0.1  &  1.6$\pm$0.1  &  93.6$\pm$0.5  &  224.9$\pm$1.4  &  4.0$\pm$0.1  &  13.1$\pm$0.1  \\
4569  &  2.0$\pm$0.1  &  116.1$\pm$0.3  &  242.4$\pm$0.9  &  4.8$\pm$0.1  &  17.2$\pm$0.1  &  1.7$\pm$0.1  &  99.8$\pm$0.6  &  206.7$\pm$1.5  &  3.8$\pm$0.1  &  14.1$\pm$0.1  \\
4570  &  1.9$\pm$0.1  &  113.7$\pm$0.3  &  252.0$\pm$0.7  &  4.7$\pm$0.1  &  17.0$\pm$0.1  &  1.6$\pm$0.1  &  94.1$\pm$0.6  &  207.0$\pm$1.4  &  3.9$\pm$0.1  &  14.1$\pm$0.1  \\
4572  &  1.9$\pm$0.1  &  109.6$\pm$0.6  &  253.1$\pm$1.3  &  4.8$\pm$0.1  &  16.2$\pm$0.2  &  1.4$\pm$0.1  &  79.1$\pm$0.7  &  181.6$\pm$1.3  &  4.6$\pm$0.1  &  12.1$\pm$0.2  \\
4573  &  1.9$\pm$0.1  &  112.5$\pm$0.3  &  240.3$\pm$1.7  &  4.8$\pm$0.1  &  16.0$\pm$0.1  &  1.5$\pm$0.1  &  90.4$\pm$0.7  &  221.4$\pm$2.4  &  3.5$\pm$0.1  &  13.3$\pm$0.2  \\
4575  &  1.4$\pm$0.1  &  77.8$\pm$0.7  &  188.5$\pm$1.1  &  5.9$\pm$0.1  &  12.1$\pm$0.2  &  1.5$\pm$0.1  &  83.5$\pm$0.9  &  203.3$\pm$1.6  &  5.8$\pm$0.1  &  13.2$\pm$0.3  \\
4581  &  1.9$\pm$0.1  &  111.3$\pm$0.3  &  247.4$\pm$0.9  &  4.7$\pm$0.1  &  16.1$\pm$0.1  &  1.3$\pm$0.1  &  76.6$\pm$0.6  &  170.1$\pm$1.6  &  3.4$\pm$0.1  &  11.0$\pm$0.1  \\
4582  &  1.9$\pm$0.1  &  107.2$\pm$0.3  &  241.1$\pm$0.8  &  4.8$\pm$0.1  &  17.2$\pm$0.1  &  1.6$\pm$0.1  &  90.5$\pm$0.5  &  201.5$\pm$1.2  &  4.4$\pm$0.1  &  14.1$\pm$0.1  \\
4583  &  1.9$\pm$0.1  &  107.1$\pm$0.4  &  228.1$\pm$1.2  &  4.8$\pm$0.1  &  17.3$\pm$0.1  &  1.6$\pm$0.1  &  90.6$\pm$0.6  &  193.4$\pm$1.6  &  4.5$\pm$0.1  &  14.1$\pm$0.2  \\
4584  &  1.8$\pm$0.1  &  100.2$\pm$0.7  &  246.7$\pm$1.5  &  4.9$\pm$0.1  &  17.1$\pm$0.2  &  0.5$\pm$0.1  &  31.1$\pm$0.5  &  74.4$\pm$0.9  &  4.0$\pm$0.1  &  5.2$\pm$0.1  \\
4585  &  1.8$\pm$0.1  &  97.7$\pm$0.6  &  225.3$\pm$1.7  &  4.8$\pm$0.1  &  16.1$\pm$0.2  &  1.4$\pm$0.1  &  76.0$\pm$0.8  &  179.3$\pm$2.1  &  4.6$\pm$0.1  &  13.2$\pm$0.2  \\
4587  &  1.9$\pm$0.1  &  110.8$\pm$0.3  &  270.9$\pm$0.8  &  5.0$\pm$0.1  &  18.1$\pm$0.1  &  1.6$\pm$0.1  &  90.4$\pm$0.5  &  224.8$\pm$1.4  &  4.4$\pm$0.1  &  14.3$\pm$0.1  \\
4588  &  1.9$\pm$0.1  &  108.4$\pm$0.3  &  243.6$\pm$0.7  &  4.9$\pm$0.1  &  17.3$\pm$0.1  &  1.6$\pm$0.1  &  88.2$\pm$0.7  &  203.7$\pm$1.4  &  4.1$\pm$0.1  &  14.4$\pm$0.1  \\
4589  &  1.8$\pm$0.1  &  103.4$\pm$0.3  &  248.9$\pm$0.8  &  4.7$\pm$0.1  &  16.1$\pm$0.1  &  1.6$\pm$0.1  &  91.2$\pm$0.6  &  219.6$\pm$1.4  &  4.0$\pm$0.1  &  14.1$\pm$0.2  \\
4590  &  1.9$\pm$0.1  &  107.4$\pm$0.5  &  260.1$\pm$1.2  &  4.9$\pm$0.1  &  17.3$\pm$0.1  &  1.4$\pm$0.1  &  77.2$\pm$0.6  &  188.2$\pm$1.4  &  4.6$\pm$0.1  &  12.1$\pm$0.1  \\
4591  &  1.9$\pm$0.1  &  106.5$\pm$0.6  &  249.8$\pm$1.4  &  4.6$\pm$0.1  &  15.1$\pm$0.1  &  1.5$\pm$0.1  &  84.6$\pm$0.8  &  199.4$\pm$1.9  &  4.3$\pm$0.1  &  12.2$\pm$0.2  \\
4592  &  2.0$\pm$0.1  &  112.6$\pm$0.5  &  290.8$\pm$0.8  &  4.9$\pm$0.1  &  17.3$\pm$0.1  &  1.5$\pm$0.1  &  84.2$\pm$0.8  &  214.9$\pm$1.2  &  4.5$\pm$0.1  &  13.1$\pm$0.2  \\
4593  &  2.0$\pm$0.1  &  109.2$\pm$0.5  &  265.0$\pm$1.3  &  4.8$\pm$0.1  &  17.0$\pm$0.1  &  1.3$\pm$0.1  &  69.4$\pm$0.8  &  172.4$\pm$1.6  &  4.6$\pm$0.1  &  11.2$\pm$0.2  \\
4594  &  2.1$\pm$0.1  &  108.2$\pm$1.1  &  266.9$\pm$2.1  &  4.9$\pm$0.1  &  17.5$\pm$0.3  &  1.5$\pm$0.1  &  78.2$\pm$0.7  &  191.0$\pm$1.6  &  4.9$\pm$0.1  &  12.1$\pm$0.2  \\
4595  &  2.5$\pm$0.1  &  103.4$\pm$0.8  &  268.7$\pm$1.9  &  5.1$\pm$0.1  &  18.1$\pm$0.2  &  0.5$\pm$0.1  &  21.2$\pm$0.5  &  56.9$\pm$0.8  &  3.9$\pm$0.1  &  3.7$\pm$0.1  \\
4596  &  2.1$\pm$0.1  &  119.1$\pm$0.5  &  277.9$\pm$1.2  &  5.0$\pm$0.1  &  18.6$\pm$0.1  &  1.6$\pm$0.1  &  87.2$\pm$0.8  &  205.5$\pm$1.7  &  4.4$\pm$0.1  &  13.1$\pm$0.2  \\
4597  &  1.9$\pm$0.1  &  107.5$\pm$0.7  &  262.4$\pm$1.4  &  4.9$\pm$0.1  &  17.2$\pm$0.2  &  1.6$\pm$0.1  &  89.2$\pm$0.8  &  217.6$\pm$1.5  &  4.8$\pm$0.1  &  14.2$\pm$0.2  \\
4598  &  2.1$\pm$0.1  &  107.1$\pm$1.2  &  256.7$\pm$2.2  &  4.7$\pm$0.1  &  16.3$\pm$0.3  &  1.2$\pm$0.1  &  60.8$\pm$1.1  &  147.3$\pm$1.6  &  4.6$\pm$0.1  &  9.5$\pm$0.2  \\
4600  &  2.0$\pm$0.1  &  111.7$\pm$0.5  &  279.3$\pm$0.8  &  4.7$\pm$0.1  &  16.2$\pm$0.1  &  1.6$\pm$0.1  &  88.1$\pm$1.1  &  225.0$\pm$1.3  &  4.4$\pm$0.1  &  13.1$\pm$0.2  \\
4601  &  2.1$\pm$0.1  &  95.8$\pm$2.1  &  251.8$\pm$6.7  &  4.7$\pm$0.1  &  17.3$\pm$0.3  &  1.4$\pm$0.1  &  69.9$\pm$1.9  &  212.7$\pm$3.7  &  4.1$\pm$0.1  &  11.2$\pm$0.4  \\
4602  &  1.9$\pm$0.1  &  106.4$\pm$0.5  &  242.8$\pm$0.9  &  4.8$\pm$0.1  &  17.1$\pm$0.1  &  1.7$\pm$0.1  &  93.8$\pm$0.7  &  211.6$\pm$1.5  &  4.6$\pm$0.1  &  15.3$\pm$0.2  \\
4603  &  2.0$\pm$0.1  &  109.0$\pm$0.4  &  275.9$\pm$0.8  &  4.7$\pm$0.1  &  16.6$\pm$0.1  &  1.3$\pm$0.1  &  70.3$\pm$0.6  &  178.0$\pm$1.2  &  4.2$\pm$0.1  &  10.6$\pm$0.1  \\
4604  &  1.9$\pm$0.1  &  110.1$\pm$0.4  &  275.1$\pm$1.1  &  4.7$\pm$0.1  &  17.8$\pm$0.1  &  1.5$\pm$0.1  &  85.2$\pm$0.9  &  220.2$\pm$1.7  &  4.3$\pm$0.1  &  13.7$\pm$0.2  \\
4605  &  1.9$\pm$0.1  &  107.6$\pm$0.5  &  291.8$\pm$0.9  &  4.7$\pm$0.1  &  16.9$\pm$0.1  &  1.7$\pm$0.1  &  96.7$\pm$0.8  &  261.3$\pm$1.4  &  4.6$\pm$0.1  &  15.3$\pm$0.2  \\
4607  &  2.1$\pm$0.1  &  109.0$\pm$0.8  &  293.9$\pm$1.3  &  5.0$\pm$0.1  &  17.4$\pm$0.2  &  1.5$\pm$0.1  &  76.6$\pm$1.7  &  211.3$\pm$2.1  &  4.5$\pm$0.1  &  12.5$\pm$0.4  \\
4608  &  2.2$\pm$0.1  &  113.0$\pm$0.5  &  253.5$\pm$2.1  &  4.8$\pm$0.1  &  17.9$\pm$0.1  &  1.6$\pm$0.1  &  81.3$\pm$1.1  &  191.3$\pm$3.2  &  4.3$\pm$0.1  &  12.2$\pm$0.3  \\
4613  &  2.1$\pm$0.1  &  106.7$\pm$1.3  &  246.9$\pm$1.9  &  5.0$\pm$0.1  &  17.6$\pm$0.3  &  1.5$\pm$0.1  &  80.6$\pm$1.4  &  176.5$\pm$2.3  &  4.9$\pm$0.1  &  12.4$\pm$0.4  \\
4614  &  1.9$\pm$0.1  &  102.8$\pm$0.9  &  219.7$\pm$1.7  &  4.9$\pm$0.1  &  15.7$\pm$0.2  &  1.6$\pm$0.1  &  88.4$\pm$1.1  &  188.0$\pm$1.8  &  4.8$\pm$0.1  &  13.2$\pm$0.3  \\
4615  &  2.4$\pm$0.1  &  128.2$\pm$0.6  &  316.5$\pm$1.4  &  4.9$\pm$0.1  &  18.1$\pm$0.1  &  2.5$\pm$0.1  &  130.2$\pm$2.1  &  342.6$\pm$3.2  &  4.2$\pm$0.1  &  19.8$\pm$0.5  \\
4616  &  2.2$\pm$0.1  &  116.3$\pm$0.6  &  244.9$\pm$1.1  &  4.9$\pm$0.1  &  17.6$\pm$0.1  &  1.7$\pm$0.1  &  91.1$\pm$0.6  &  190.8$\pm$1.2  &  4.8$\pm$0.1  &  13.2$\pm$0.2  \\
4617  &  1.9$\pm$0.1  &  110.1$\pm$0.8  &  274.8$\pm$2.1  &  5.0$\pm$0.1  &  17.6$\pm$0.2  &  8.8$\pm$0.1  &  482.6$\pm$6.1  &  1230.3$\pm$12.3  &  4.9$\pm$0.1  &  75.9$\pm$1.5  \\
4618  &  1.8$\pm$0.1  &  117.5$\pm$2.4  &    &  5.5$\pm$0.1  &  19.4$\pm$0.5  &  1.2$\pm$0.1  &  72.6$\pm$3.7  &  142.2$\pm$3.5  &  4.1$\pm$0.2  &  12.3$\pm$0.7  \\
\hline
\end{tabular}\\
Note: the first column shows the HJD information, the Second to the 
sixth columns show the parameters from the scaled spectra: the continuum 
emission at 5100\AA\ in unit of $10^{-15}erg/s/cm^2/\AA$, the line fluxes 
of broad H$\beta$ ($f_\beta$) and broad H$\alpha$ ($f_\alpha$) 
in unit of $10^{-15}erg/s/cm^2$, the second moment of the \oiii line 
($\sigma_3$) in unit of \AA, and the line flux of \oiii ($f_3$) in 
unit of $10^{-15}erg/s/cm^2$. The last five columns shows the parameters 
from the final spectra.  
\end{minipage}
\end{table*}

\section{Conclusions}
   Finally, our main conclusions are as follows. On the one hand, 
we carefully checked whether the spectral scaling calibration method is 
preferred for the LAMP final spectra of Mrk142, and we can find that 
with the application of the spectral scaling calibration method based on 
the assumption of constant \oiii line, we do not find any stronger parameter 
correlations. Therefore, the commonly applied spectral 
scaling calibration method, based on the assumption of a constant \oiii line 
(especially constant \oiii line flux) is not preferred for the spectra of 
Mrk142. On the other hand, we find there is reliable variability 
of the \oiii line, based on the strong correlations between \oiii line 
flux and continuum emission, and between \oiii line flux and broad balmer 
line flux.  

\section*{Acknowledgements}
Zhang and FLL gratefully acknowledge the anonymous referee for giving us 
constructive comments and suggestions to greatly improve our paper. 
Zhang acknowledges the kind support from the Chinese grant NSFC-U1431229. 
FLL is supported under the NSFC grants 11273060, 91230115 and 11333008, 
and State Key Development Program for Basic Research of China 
(No. 2013CB834900 and 2015CB857000). This work has made use of data from 
the Lick AGN Monitoring Project public data release.

\label{lastpage}
\end{document}